\documentclass[preprint,showpacs,preprintnumbers,aps,pop]{revtex4-1}

\usepackage{graphicx}
\usepackage{bm}
\usepackage{amsmath}
\usepackage{amssymb}
\usepackage[usenames]{color}

\newcommand{\dede}[2]{\frac{\delta #1}{\delta #2}}

\newcommand{\rem}[1]{}
\newcommand{\de}{{\rm d}}

\newcommand{\bq}{{\mathbf{x}}}
\newcommand{\bv}{{\mathbf{v}}}

\newcommand{\bM}{{\mathbf{M}}}

\newcommand{\bB}{{\mathbf{B}}}

\newcommand{\bU}{{\boldsymbol{U}}}

\newcommand{\hod}{\hat{z}\times\nabla}
\newcommand{\nablaperp}{\nabla_{\perp}}
\newcommand{\mf}{\mathcal{F}}

\newsavebox{\fminibox}
\newlength{\fminilength}

\begin{document}

\title{Energy stability analysis for a hybrid fluid-kinetic plasma model}
\author{Philip J. Morrison}
\email{morrison@physics.utexas.edu}
\affiliation{Department of Physics \& Institute for Fusion Studies, University of Texas, Austin 78712-0262, USA}
 \author{Emanuele Tassi}
\email{emanuele.tassi@cpt.univ-mrs.fr}
\affiliation{CNRS \& Centre de Physique Th\'eorique, Campus de Luminy, 
13288 Marseille cedex 9, France and Universit\'e de Toulon, CNRS, CPT, UMR 7332, 83957, La Garde, France}
\author{Cesare Tronci}
\email{c.tronci@surrey.ac.uk}
\affiliation{Department of Mathematics, University of Surrey, Guildford GU2 7XH, United Kingdom}

\begin{abstract}
In plasma physics, a hybrid fluid-kinetic model is composed of a magnetohydrodynamics (MHD) part that describes a bulk fluid component and a Vlasov kinetic theory part that describes an energetic plasma component. While most hybrid models in the plasma literature are non-Hamiltonian, this paper investigates a recent Hamiltonian variant in its two-dimensional configuration. The corresponding Hamiltonian structure is described along with its Casimir invariants. 
Then, the energy-Casimir method is used to derive explicit sufficient stability conditions,  which imply a stable spectrum and suggest nonlinear stability. 
\end{abstract}

\maketitle


\section{Introduction}

The discipline of plasma physics has  provided a rich collection of spectral and stability problems.  
Depending on the circumstances, plasmas may be described by fluid models or kinetic theories with coupling to electromagnetic fields.  Consequently,  plasma theory has all of the possibilities and concomitant complications of all these disciplines and, although a great deal of lore has been generated on the formal level, there remain many open mathematical problems of a spectral and stability nature.

When dissipative terms are small enough to be safely neglected, the resulting  plasma systems should be infinite-dimensional Hamiltonian systems.  Thus, one is led to  stability and spectral problems of a wide variety of Hamiltonian operators. Usually these operators are non-Hermitian, non-normal, and have rich spectra with both point and continuous components.  Bifurcation theory of these operators including their nonlinear extensions provide  challenging nontrivial problems of important physical relevance. 

A particularly challenging class of spectral and stability  problems arises from the so-called hybrid models of plasma physics.  These are models the incorporate both fluid and kinetic equations.   Generally speaking, the purpose of these models is to describe  a bulk portion of the plasma by a fluid model, such as MHD,   while describing a hot component of the plasma by a kinetic theory, such as the Vlasov-Maxwell system.  Thus, hybrid models can combine all spectral and stability issues that occur in fluid and kinetic theories separately into a complicated whole. It is well-known that MHD and the Vlasov equation separately have a variety of interesting point and continuous components to their spectra and so hybrid models can indeed present a challenging class of mathematical problems.  However,  because of the Hamiltonian nature of these models one can use energy techniques to obtain spectral information without performing detailed operator analysis.  In particular, one can obtain sufficient conditions for the existence of a stable spectrum by the so-called energy-Casimir method (see \cite{HoMaRaWe,Mor98}) that is based on the natural Lyapunov-Dirichlet method of Hamiltonian systems theory.  

The purpose of this article is to describe a particular hybrid model that is a coupling between two-dimensional (planar)  MHD and Vlasov theory.  We will describe its Hamiltonian structure and apply the energy Casimir method to a class of equilibrium states and obtain sufficient conditions for stability.   The article is organized as follows.  In Sec.~\ref{sec:CS}  we review some details regarding stability and the energy-Casimir method.  Then in Sec.~\ref{sec:planhybrid} we describe the planar hybrid model, its noncanonical Hamiltonian structure, and associated Casimir invariants.  This is followed in Sec.~\ref{sec:enCas} by the application of the energy-Casimir method, giving rise to the sufficient conditions.  Finally, we conclude in  Sec.~\ref{sec:conclu}.  For completeness, two appendices  are included:   Appendix \ref{app1}  provides a proof of the Jacobi identity, while Appendix \ref{app2} contains a  direct verification that the functional indicated as Casimir for the hybrid model, indeed commutes with every other observable.


\section{Stability and the energy-Casimir method}    
\label{sec:CS}

Consider a dynamical system $\dot z= V$ defined on some space (manifold)  $\mathcal{Z}$, 
where $z\in\mathcal{Z}$ can be a point or a trajectory, `$\ \dot{\ }\ $' denotes time derivative,  and $V$ is an autonomous 
vector field defined on $\mathcal{Z}$.  Our interest is in the stability of equilibrium points $z_e$, solutions that satisfy $V(z_e)=0$.   We adopt the standard definition  of stability for such points, which  is the following:
\begin{quote}
An equilibrium point $z_e$ of a dynamical system is said to be stable if, for any neighborhood $N$ of $z_e$ there exists a subneighborhood $S\subset N$ of $z_e$ such that if $\mathring{z}$, an initial condition,  is in $S$ then the trajectory $z(t)\in N$ for all time $t>0$.
\end{quote}
 
\vspace{-.2 cm} \noindent Often one considers the associated linear problem $\delta\dot{z}= DV(z_e)\cdot \delta z$  obtained by expanding $V(z_e + \delta z)$ to first order.  If  $\delta z$ remains in $N$, then the  system is said to be linearly stable, and to  distinguish this kind of stability from that with dynamics under the full vector field $V$ one adds the adjective nonlinear to describe the latter. Assuming a solution of the form $\delta z= \hat{z}\exp(\lambda t)$ the linear problem becomes $(DV-\lambda \, {\rm id})\cdot \hat{z}=0$, where id is the identity operator.  The spectrum of $DV$, 
$\sigma(DV)$,  is the set composed of   $\lambda\in \mathbb{C}$ for which the linear operator $DV-\lambda \, {\rm id}$ 
has no inverse, and an equilibrium point is said to be spectrally stable if $i\sigma(DV)\subset \overline{{\rm LHP}}$.  Observer that this definition of stability includes the case $i\lambda \in \mathbb{R}$, which corresponds to pure oscillation, a case that is sometimes called neutral stability. Inclusion of this case is most important, since this is the only kind of spectral stability possessed by Hamiltonian systems. 

There are several  logical implications between the different types of stability, and these can be  somewhat subtle: for example, linear stability implies spectral stability;  linear stability does not imply nonlinear stability; nonlinear stability does not imply linear stability.  For establishing  linear stability in Hamiltonian systems, which would assure us that the spectrum lines on $i\mathbb{R}$,  one can use Lyapunov function techniques dating back to Lagrange and Dirichlet.  Our main concern of this paper is to do this for a particular hybrid model.   

For Hamiltonian systems, the vector field of our dynamical system is generated by a Poisson bracket so that the equations of motion have the form $\dot{z}=\{z,H\}$ which for finite-dimensional systems in a coordinate patch is given by $\dot{z}^i=J^{ij}\partial_jH$, where $J$ is the Poisson bivector (cosymplectic form) and $\{\ ,\ \}\colon C^{\infty}(\mathbb{R})\times C^{\infty}(\mathbb{R})\rightarrow  C^{\infty}(\mathbb{R)}$.  Thus equilibrium points satisfy $dH\in {\rm Ker}(J)$.  A consequence of the Poisson bracket identities (see Appendix \ref{app1}) is the Lie-Darboux theorem, which implies  ${\rm Ker}(J)$ is spanned by Casimir invariants, which satisfy $\{C,f\}=0$, for all functions $f\in C^{\infty}(\mathbb{R})$ (although there are serious unresolved issues with this theorem for infinite-dimensional systems (see, e.g., \cite{yoshida}).  Thus equilibria are critical points of an invariant energy function $\mathcal{F}= H + C$ and this fact is very useful for establishing stability criteria. 

For finite-dimensional systems, definiteness (positive or negative) of the quadratic form $\delta^2\mathcal{F}(z_e;\delta z)$ assures both linear and nonlinear stability.  For linear systems $\delta^2\mathcal{F}$ is invariant, in fact the Hamiltonian for the linear dynamics,  and its definiteness means the equilibrium point is foliated by nested invariants sets that are topologically spheres. The interior of these sets can thus serve as the subneighborhoods $S$ in the above definition of stability.  For nonlinear systems $\mathcal{F}$ is invariant and, under mild smoothness conditions, $\delta^2\mathcal{F}$ determines the topological character of the level sets of $\mathcal{F}$ in a sufficiently small neighborhood of the equilibrium point $z_e$.  This guarantees that for any neighborhood $N$ there is an $S\subset N$,  determined by some level set of $\mathcal{F}$, within which the flow must remain.  

For infinite-dimensional systems, the situation is considerably more complicated.  First, definiteness of 
$\delta^2\mathcal{F}$ does not imply that an extremal point that satisfies $\delta\mathcal{F}=0$ is in fact an extermum (maximum or minimum).   For both linear and nonlinear stability one requires $\delta^2\mathcal{F}$ to lead to a norm for defining open sets.  This leads to a second and for nonlinear stability an often formidable complication:  even if a norm can be extracted from $\delta^2\mathcal{F}$ for a rigorous proof of nonlinear stability one must show that the solution to the dynamical system actually exists in this norm. Unfortunately, for the systems of most physical interest in plasma physics, global existence results are not available.  In the rosy situation where existence results are firm, it  can  turn out that  more than one choice of norm may be available   and a given  equilibrium  can be stable in one norm and not another;  in this  case a physical determination must be made about what is important.  

In this paper we will follow the practice in the physics literature (e.g.\  \cite{Haz84,HoMaRaWe,MorEli,Mor98})  and only obtain  formal stability criteria for our hybrid models.  More specifically, in Sec.~\ref{sec:enCas} we will find conditions under which $\delta^2\mathcal{F}$ is positive definite.  The reader interested in seeing what makes up a rigorous application of the energy-Casimir method is referred to Refs.~\cite{Batt,Rein}.


\section{Planar Hamiltonian hybrid model}
\label{sec:planhybrid}

 Many hybrid models exist in the plasma physics literature, but a most popular kinetic-MHD variant is that of \cite{Cheng,ParkEtAl,PaBeFuTaStSu,FuPark}, which has been  used  often in computer simulations  \cite{KimEtAl,TaBrKi}. This model employs  the so-called pressure-coupling scheme, which suffers from not conserving energy exactly. Recently, a Hamiltonian version of this scheme (HPCS) was given in \cite{HoTr2011,Tronci,TrTaMo}).
 Here we will present and analyze a two-dimensional variant of the HPCS.  The equations of motion will be given in Sec.~\ref{ssec:HMeom}, its Hamiltonian structure in  Sec.~\ref{ssec:hamstru}, and its Casimir invariants in Sec.~\ref{ssec:cassies}.

 \subsection{Planar hybrid model equations of motion}
 \label{ssec:HMeom}

Upon setting all physical constants equal to unity, the planar Hamiltonian hybrid model is given by the following system of partial differential equations: 
\begin{align}
&\partial_t\omega=\left[\psi,\omega\right]+\left[J,A\right] 
+\hod\cdot\left(\nablaperp\cdot\!\int\!\de^3v\, f\,  \bv_\perp\bv_\perp \right)\,, \label{vort}
\\
&\partial_t A=[\psi,A] \,,  \label{ohm}
\\
&\frac{\partial f}{\partial t}+\left[f,\psi\right]+\bv_\perp\cdot\frac{\partial f}{\partial\bq_\perp}+\bv\cdot\hod \left(\nablaperp \psi \cdot \frac{\partial f}{\partial \bv_\perp} \right) \nonumber
\\
&\hspace{3 cm} +v_z\nablaperp A \cdot\frac{\partial f}{\partial \bv_\perp} 
 -\bv_\perp\cdot\nablaperp A\,\frac{\partial f}{\partial v_z}=0 \,.
 \label{vla}
\end{align}
In these equations, the scalar functions $A$ and $\omega$ are defined on a domain $\mathcal{D} \subseteq \mathbb{R}^2$ and indicate the magnetic poloidal flux function and the vorticity of the bulk flow. These are related to the magnetic field $\bB$ and to the bulk velocity field $\bU$ by 
\[
\omega=\hod \cdot \bU
\,,\qquad\ 
\bB=-\hod A,
\]
 $\hat{z}$ indicates the unit vector along the coordinate $z$ of a Cartesian system $(x,y,z)$,  in which the coordinates $x$ and $y$ cover the domain $\mathcal{D}$. The two-dimensional  gradient $\nablaperp$ acts as $\nablaperp u = \hat{x} \partial_x u + \hat{y} \partial_y u$ on a generic function $u$. The  current density $J$ and the stream function $\psi$ , on the other hand, are related to the magnetic flux function and to the vorticity by
\begin{equation}
J=-\Delta A, \qquad \psi=-\Delta^{-1} \omega,
\label{constraints}
\end{equation}
respectively.  In (\ref{constraints}), the symbol  $\Delta$ denotes the two-dimensional Laplacian.  The distribution function (phase space density) $f(\bq_\perp, \bv)$ is defined over the  particle phase space $\mathcal{D} \times \mathbb{R}^3$, where $\bq_\perp$ and $\bv_\perp$ denote $(x,y)$ and $(v_x , v_y)$, respectively. Finally, we indicated by $[\,  ,\,  ]$ the canonical bracket acting on two functions as $f$ and $g$ by $[f,g]:=\nablaperp g \cdot \hod f$.
	
Equations (\ref{vort})--(\ref{vla}) govern the evolution of an incompressible MHD bulk system, coupled with a kinetic particle population. Equation (\ref{vort}) is a vorticity equation, in which the bulk plasma flow is affected by the presence of the kinetic species, through the additional pressure divergence term, represented by the last term on the right-hand side of equation  \eqref{vort}. In the absence of such term, one retrieves the two-dimensional  version of classical MHD. Analogously, Eq.~(\ref{vla}), describes the evolution of the distribution function of the kinetic species, which, in its turn, is influenced by the transport and force term associated with the bulk velocity. Such effects vanish upon setting $\psi=0$ in (\ref{vla}). Equation  (\ref{ohm}), on the other hand, is the  ideal Ohm's law reflecting the assumption that the magnetic flux  is frozen into the bulk fluid.


\subsection{Hamiltonian structure}
\label{ssec:hamstru}

The  model of  Eqs.~(\ref{vort})--(\ref{vla}) is  easily obtained from the three-dimensional  Hamiltonian hybrid model   in \cite{Tronci,TrTaMo} by reduction to two spatial dimensions.  Thus, it is not surprising that it inherits a Hamiltonian structure in terms of a noncanonical Poisson bracket.  In Appendix \ref{app1} we show how to reduce the Poisson bracket of the three-dimensional model to obtain  the planar system \eqref{vort}-\eqref{vla}.

Recall, a  Poisson bracket  $\{ \,,\, \}$ defines a Lie algebra realization on a set  of  observables consisting of the functionals of the dynamic variables, which here are  $\omega$, $A$, and $f$.  As alluded to in Sec.~\ref{sec:CS},  time evolution of an element $F$ of such algebra is determined 
by the equation
\begin{equation}  \label{heqs}
\partial_t F=\{F,H\},
\end{equation}
where, for the case at hand, the Hamiltonian $H$ is given by
\begin{equation}
H=\frac12\int\! \de^2x \Big(-\omega\Delta^{-1}\omega-A\,\Delta A+\int \! \de^3v\, f|\bv|^2\Big)\,,
 \label{ham}
\end{equation}
and the expression for the Poisson bracket reads
\begin{align}\nonumber
\{F,G\}=&\int \de^2 x \, \omega\left[\dede{F}{\omega},\dede{G}{\omega}\right]+\int \de^2 x \, A\left(\left[\dede{F}{\omega},\dede{G}{A}\right]-\left[\dede{G}{\omega},\dede{F}{A}\right]\right)
\\\nonumber
&-\int \de^2 x  \de^3 v \, f\left(\left[\dede{F}{f},\bv \cdot \hod \dede{G}{\omega}\right]_v 
-\left[\dede{G}{f},\bv\cdot\hod\dede{F}{\omega}\right]_v\right)
\\
& \hspace{.3 cm}+\int\! \de^2 x \de^3 v \, f\left(\left[\dede{F}{f},\dede{G}{f}\right]_v 
\right.
\label{planarbracket}\\
&\left.\hspace{.75 cm} + \, \nablaperp A\cdot\!\left(\frac{\partial}{\partial v_z}\dede{G}{f}\frac{\partial}{\partial \bv_\perp}\dede{F}{f}-\frac{\partial}{\partial v_z}\dede{F}{f}\frac{\partial}{\partial \bv_\perp}\dede{G}{f}\right)\right).
\nonumber
\end{align}
	This bilinear operation satisfies antisymmetry, the Leibniz identity and the Jacobi identity (cf.\ Appendix \ref{app1}). In (\ref{planarbracket}) we introduced a canonical bracket defined over a reduced phase space as $[f,g]_v := \nablaperp f \cdot \partial_{\bv_\perp}g - \nablaperp g \cdot \partial_{\bv_\perp} f$.

For the choices $F=\omega$, $A$,  or $f$, using (\ref{ham}) and (\ref{planarbracket}) in (\ref{heqs}), one retrieves the model equations (\ref{vort})--(\ref{vla}), provided that boundary terms arising from integration by parts vanish. This is accomplished, for instance, if the involved functions are periodic on $\mathcal{D}$, or, in case $\mathcal{D}$ is unbounded, if they also decay at infinity. The functions depending on the velocity coordinates are also assumed to go to zero sufficiently fast as $\bv \rightarrow \infty$.

Concerning the Hamiltonian (\ref{ham}), we remark that it naturally expresses the total energy of the system, consisting of the sum of the bulk kinetic energy, the magnetic energy and the kinetic energy of the hot particle population, corresponding to the three terms appearing in (\ref{ham}), respectively.

With regard to the Poisson bracket, on the other hand, we observe that it possesses a pure MHD part, consisting of the first two terms of (\ref{planarbracket}), which correspond to the Poisson bracket of reduced MHD \cite{Mor84,MaMo}. It possesses also a purely kinetic part, given by its last two terms, which include the Vlasov bracket \cite{Morrison2bis,Morrison3,MaWe1}. The remaining   terms, on the other hand, are those responsible for the coupling between the MHD and the kinetic components.


\subsection{Casimir invariants}
\label{ssec:cassies}

As anticipated in Sec.~\ref{sec:CS}, the energy-Casimir method requires the identification of the Casimir invariants $C$, i.e.,  functionals satisfying $\{F,C\}=0$ for any arbitrary functional $F$ in the algebra of observables. However, finding the Casimirs is not always an easy task and their limited availability stands sometimes as the major obstacle to the application of the method. However, in the case under consideration, the existence of a cross-helicity invariant for the Hamiltonian PCS was shown in \cite{HoTr2011} and this Casimir finds its way into the present two-dimensional theory, where it is generalized.  The existence of such a Casimir for the three-dimensional Poisson bracket yields a whole family of Casimir invariants when projected on the plane. Upon summing contributions arising from the magnetic helicity and the Vlasov dynamics, the total Casimir invariant for the planar hybrid model reads 
\begin{equation}\label{NewCasimir}
C(\omega, A, f)=\int \de^2 x\, 
 \left(
\overline{\omega}\, \Phi(A)
 +\Psi(A)+\int \de^3 v\,  \Lambda(f)\right),
\end{equation}
where   $\Phi$, $\Psi$,  and $\Lambda$ are arbitrary functions,  and we have introduced the shorthand 
\[
\overline{\omega}:=\omega -\hod\cdot  \mathbf{K}\,,\qquad  \textrm{with}\qquad \mathbf{K}:=\int\! \de^3v\, f\, \bv_\perp
\,.
\] 
That (\ref{NewCasimir}) is in fact a Casimir invariant is shown in Appendix \ref{app2}.

The existence of such Casimir invariants is amenable to a physical interpretation, which becomes clearer when considering separately, the contributions coming from the  functions $\Phi$, $\Psi$,  and $\Lambda$. If $\Phi\equiv\Lambda\equiv 0$, the remaining Casimir expresses the conservation of magnetic flux through a surface moving with the bulk fluid velocity. This property is the frozen-in condition for the magnetic flux, inherited from ideal MHD. If $\Phi\equiv\Psi\equiv 0$ one retrieves the conservation of the integral of any function of $f$, which is characteristic of Vlasov systems and whose physical meaning in terms of particle rearrangements was  given in  \cite{Mor87}. On the other hand, a new family of invariants  associated with this  two-dimensional hybrid model appears when setting $\Psi\equiv\Lambda\equiv 0$. For this case,  the Casimir family reduces to
\begin{eqnarray}
C(\omega, A, f)&=&\int \de^2 x \,  \left(\omega - \hod \cdot\int\! \de^3 v\, f\,  \bv_\perp \right)\Phi(A)
 \nonumber\\
&=&\int \de^2 x \,  \Phi ' (A)\left(\nablaperp \psi \cdot \nablaperp A  
+ \int \de^3 v\,  f \, \bv_\perp \cdot \hod A\right)
 \nonumber \\
&=&\int \de^2 x\,  \Phi ' (A) (\bU - \mathbf{K}) \cdot \bB\,, 
 \label{chel}
\end{eqnarray}
where recall $\mathbf{K} = \int \de^3 v \, f\,  \bv_\perp$, which  corresponds  to  the momentum of the hot particle species in the $xy$-plane. Equation  (\ref{chel}) introduces a hybrid cross-helicity density $(\bU - \mathbf{K}) \cdot \bB$, expressing the correlation between the magnetic field and a velocity field obtained by subtracting from  the bulk velocity a contribution coming from the kinetic species.  Upon setting $\mathbf{K}\equiv 0$ this Casimir reduces to the cross helicity family of invariants for  two-dimensional MHD which to our knowledge was first found in \cite{Mor84}.  For a given constant $A_0$, choosing $\Phi(A)= H_{A_0}(A)$, with $H$ indicating the Heaviside function with step located at $A=A_0$, expresses the property that, not only is  the total generalized cross-helicity is conserved, but also its integral over domains $A=A_0$, which are bounded by magnetic flux surfaces.

Given that these families of  Casimirs have been identified explicitly, we are ready to apply the energy-Casimir method and find sufficient conditions for energy stability of the system. This is carried out in the next section.


\section{Energy-Casimir  stability analysis}
\label{sec:enCas}

Knowledge of the Casimir invariants  provides a variational principle for equilibria, $\delta \mf=0$, which we tend to in 
Sec.~\ref{ssec:equil}.  Then, the next step of the energy-Casimir method is to consider the second variation, $\delta^2 \mf=0$, which is done in Sec.~\ref{ssec:stabcon}. Notice that all physical constants have been set to unity.  A more perspicuous  study of the dynamical behavior can be obtained upon restoring these constants. 


\subsection{Equilibrium variational  principle}
\label{ssec:equil}

To construct a variational principle for the equilibria under consideration, we first define the free energy functional $\mf= H+C$.  Then,  we take  its first variation, which reads
\[
\delta \mf=\delta \mf_{MHD}+\delta \mf_V -\delta\!\iint \de^2 x \de^3 v \, \Phi(A)  \bv\cdot \hod f
\,,
\]
where we split-off the Vlasov part 
$\delta \mf_V:= \int\! \de^2 x \de^3 v \,  \left(\Lambda'+|\bv|^2/2\right) \delta f$ and the MHD part
\[
\delta \mf_{MHD}:=\int\! \de^2 x \, 
 \big(\Phi(A)- \Delta^{-1}\omega\big) \delta\omega
 +\int\! \de^2 x \, \big(\omega\Phi'(A)-\Delta A+\Psi'(A)\big)\delta A\,. 
\]
Upon setting $\delta \mf=0$, the equilibrium equations turn out to be
\begin{eqnarray}
0&=&\psi_e+\Phi(A_e)\, ,   \label{eq1}
\\
0&=&-\Delta A_e+\overline{\omega}_e\Phi'(A_e)+\Psi'(A_e)\, ,  \label{eq2}
\\
0&=&\Phi'(A_e)\bv \cdot \hod A_e+\frac{\left|\bv\right|^2}{2}+\Lambda'(f_e)   \label{eq3}
\,,
\end{eqnarray}
whose first   relation renders  the third in the form
\begin{align}
\frac12\left|\bv+\bU_e\right|^2-\frac12\left|\bU_e\right|^2+\Lambda'
=&0  \,.  \label{elf}
\end{align}
In the above expressions we introduced the subscript $e$ to indicate equilibrium quantities.

Upon assuming an invertible $\Lambda$, from (\ref{elf}) we obtain the equilibrium distribution function in the form
\begin{align}
f_e=&f_e\!\left(\frac12\left|  \bv+\bU_e\right|^2-\frac12\left|\bU_e\right|^2\right)
=
f_e\!\left(\frac12\left|\bv+\hod\Phi \right|^2-\frac12\left|\nablaperp\Phi\right|^2\right)
\,. 
\label{Vlasovequilibrium}
\end{align}
As an example, consider the relative Gaussian distribution
\[
f_e=\exp\!\left(-\frac1{2}\left|\bv+\bU_e\right|^2+\frac12\left|\bU_e\right|^2\right)
\,.
\]
It is easy to see that this yields
\begin{equation*}
\int\!  \de^3 v \, f_e\, \bv =-e^{-\frac12\left|\bU_e\right|^2}\bU_e
\quad \Longrightarrow\quad \overline{\omega}
=
\nablaperp\cdot\left(\left(1+e^{-\frac12\left|\nablaperp\Phi\right|^2}\right)\nablaperp\Phi\right)
\end{equation*}
and therefore, in the absence of MHD equilibrium flow, one has $\overline{\omega}_e=0$, which means that the vorticity ${\hod\cdot\int\!d^3v\, f\, \bv_\perp }$ associated with the hot particle flow, is also zero. In the general case of an arbitrary equilibrium of the type \eqref{Vlasovequilibrium}, this quantity is computed as
\begin{align*}
\hod \cdot\int\! \de^3 v \,f_e\, \bv_\perp&
=\int\! d^3v  \, f'_e \,    \left(\bv_\perp+\bU\right)\cdot \hod\left[\frac12\left|\bv_\perp 
+ \bU\right|^2-\frac12\left|\bU\right|^2+\frac12v_z^2\right]
\\
&=-\nablaperp\cdot\left(n_e\nablaperp\Phi\right)
\end{align*}
where $n_e=\int\! d^3 v\,   f_e\,$ is the hot particle equilibrium density. 
In conclusion, the final form of the hybrid equilibrium relation reads
\[
-\Delta A_e+\Phi'(A_e)\,\nablaperp\cdot\left((1+n_e)\nablaperp\Phi(A_e)\right)+\Psi'(A_e)=0
\,.
\]
In the absence of a hot species ($n_e\equiv 0$), this reduces to the celebrated Grad-Shafranov equation  for reduced MHD \cite{Haz84}.  Note, when $n_e\not\equiv 0$ we call the above equilibrium relation  the  \emph{hybrid Grad-Shafranov equation}. 


\subsection{Stability conditions}
\label{ssec:stabcon}

Turning now to  stability criteria , we compute the   second variation
\begin{align*}
\delta^2 \mf=&\delta^2 \mf_{MHD} + \delta^2 \mf_V- 2\iint\! \de^2 x \de^3 v \, \delta \Phi(A)\ \bv_\perp\cdot\hod \delta f \\
&\hspace{1 cm}  -\iint\! \de^2 x \de^3 v \, ( \bv_\perp\cdot \hod f ) \Phi'(A) (\delta A)^2,
\end{align*}
where we have introduced
\begin{align*}
&\delta^2 \mf_{MHD}=\int \de^2 x \, (\left|\nabla\delta\psi-\nabla\delta\Phi\right|^2 
+\left(1-(\Phi')^2\right)|\nabla\delta A|^2)
\\
&\hspace{2 cm} +\int\! \de^2 x \, \left(\omega\Phi''+\Psi''+\Phi' \Delta\Phi'\right)(\delta A)^2,
\\
&\delta^2 \mf_V=\int\! \de^2 x \de^3 v \, \Lambda''(f) (\delta f)^2,
\end{align*}
which correspond  to the second variation expressions for  reduced MHD \cite{Haz84} and  the Vlasov equation \cite{Fowler,Gar63,Mor87,HoMaRaWe,Mor90}, respectively.
After some rearrangement, the expression for the second variation, evaluated at an equilibrium solution of (\ref{eq1})--(\ref{eq3}) can be written as follows: 
\begin{align}
\delta^2 \mf (\omega_e , A_e , f_e)&=\delta^2 \mf_{MHD} (\omega_e , A_e , f_e) 
+\delta^2 \mf_V (\omega_e , A_e , f_e) 
\label{secvar}
\\
&\hspace{.75 cm}  -\iint\! \de^2 x \de^3 v \, \left(\bv \cdot \hod  f_e \right)\Phi''(A_e) (\delta A)^2 
\nonumber\\
&
\hspace{.75 cm}  +\iint\! \de^2 x \de^3 v \, f_e^{-1} \Big|\delta f 
- \delta A f_e \nablaperp \Phi' (A_e)\cdot\hat{z}\times \bv_\perp\Big|^2
 \nonumber
\\
&
\hspace{.75 cm} +\iint\! \de^2 x \de^3 v  \,  f_e^{-1}  \,\Big|\delta f 
-f_e \Phi' (A_e)\nablaperp\delta A\cdot\hat{z}\times \bv_\perp\Big|^2  
\nonumber\\
&
\hspace{.75 cm}  -2\iint \de^2 x \de^3 v \, f_e^{-1} {(\delta f)^2}  -\int \! \de^2 x \, (\operatorname{Tr}{\mathbb{P}_\perp}_e) |\nablaperp\Phi' (A_e)|^2(\delta A)^2 
\nonumber\\
&
\hspace{.75 cm} 
+ \iint\! \de^2 x \de^3 v \,  f_e\, \big(\bv_\perp\cdot\nablaperp\Phi' (A_e)\big)^2(\delta A)^2  
\nonumber \\
&
\hspace{.75 cm} -\int\! \de^2 x\,  \Phi'^2(A_e)\big (\operatorname{Tr}{\mathbb{P}_\perp}_e\big)|\nablaperp\delta A|^2
\nonumber\\
&
\hspace{.75 cm} +
\iint \! \de^2 x \de^3 v \,f_e\,  \Phi'^2(A_e)(\bv_\perp\cdot\nablaperp\delta A)^2   \nonumber
\end{align}
where we have defined ${\mathbb{P}_\perp}_e:=\int\! d^3 v\, f_e \bv_\perp\bv_\perp$, while $\operatorname{Tr}$ denotes the ordinary matrix trace.  

Because energy stability is attained (by definition) if the second variation, evaluated at the equilibrium, has a definite sign, for any perturbations $\delta A$, $\delta \omega$ and $\delta f$, we infer from (\ref{secvar}), that sufficient conditions for stability are provided by
\begin{align} 
&\vert\Phi' (A_e)\vert^2< \frac{1}{1+\operatorname{Tr}{\mathbb{P}_\perp}_e} \label{sta1}
\\
&\overline{\omega}_e\,\Phi''(A_e)+\Psi''(A_e)+\Phi' (A_e)\Delta \Phi' (A_e)
-|\nablaperp\Phi' (A_e)|^2\operatorname{Tr}{\mathbb{P}_\perp}_e
>0
\label{sta2}
\\
&\Lambda''(f_e)> 2/{f_e}\,. 
\label{sta3}
\end{align}

Notice that $\psi_e=-\Phi(A_e)$ implies $\bU_e=-\Phi'(A_e)\bB_e$, so that the   stability condition  of (\ref{sta1}) reads
\[
B_e > \left(1+\operatorname{Tr}{\mathbb{P}_\perp}_e\right)U_e\, ,   \label{cflow}
\]
where $B_e :=|\bB_e|$ and $U_e:=|\bU_e|$. 
Due to the presence of the kinetic component, our stability condition requires slower equilibrium flows, in  comparison to the corresponding condition for reduced MHD.  The latter condition \cite{Mor84}, indeed, requires MHD flows to be just sub-Alfvenic, whereas this is no longer sufficient to satisfy (\ref{sta1}) in the presence of a hot particle population.

Upon making use of $\Phi '=-\bU_e\cdot\bB_e/B_e^2=-U_e/B_e$, the condition (\ref{sta2})  can be reformulated in the following way:
\begin{eqnarray}
\frac{\bB_e \times \hat{z} \cdot \nablaperp J_e}{B_e^2}\left(1-\frac{U_e^2}{B_e^2}\right)-\frac{1}{2}\frac{\bB_e \times \hat{z}}{B_e^2}\cdot\nablaperp\left(\frac{U_e^2}{B_e^2}\right)\frac{\bB_e \times \hat{z}}{B_e^2}\cdot\nablaperp B_e^2 \label{s21} \\
+ \frac{U_e}{B_e} \frac{\bB_e \times \hat{z}}{B_e^2}\cdot \nablaperp \hod \cdot \mathbf{K}_e-\left\vert \nablaperp\left( \frac{U_e}{B_e}\right)\right\vert^2 \operatorname{Tr}{\mathbb{P}_\perp}_e>0, \label{s22}
\end{eqnarray}
which provides an  interpretation of this  stability condition in terms  of physical properties  of the  equilibrium state.  The terms of line (\ref{s21}) correspond to the same terms appearing in the energy stability condition for reduced MHD.  In particular, one can recognize in the first term  the above mentioned sub-Alfvenic condition, in addition to conditions depending on the relative direction of the equilibrium magnetic fields and the gradient of the current density.  These have been shown to be a  source for the kink and interchange instabilities observed in tokamaks in the presence of magnetic curvature \cite{Mor13}. The terms of   line (\ref{s22}) account for the new contributions due to the kinetic species. These are due to the compressibility of the hot particle equilibrium flow, and to the hot particle energy.  We observe that, in the case of static MHD equilibrium (i.e.\  $\bU_e=0$), the presence of the kinetic species has no influence on the condition (\ref{s21})--(\ref{s22}). In particular, in that limit one recovers the pure reduced MHD condition \cite{Haz84},  $\bB_e \times \hat{z} \cdot \nablaperp J_e >0$, which corresponds to a current density profile monotonically decreasing with the equilibrium flux function $A_e$.
    
Finally, upon differentiating the equilibrium relation for $\Lambda'$ with respect to $\bv$, we obtain
\[
\Lambda'' (f_e)=-\frac1{f'_e}\ \Longrightarrow\ f_e<-2f'_e\,. 
\,
\]
 For example, the particular Gaussian distributions  such that $f_e'=-f_e$ (unit variance) are stable equilibria.  This is a modification of Gardner's  well-known monotonicity theorem \cite{Gar63,Fowler}. 

It should be emphasized that our sufficient conditions are not optimal.  Clearly some potentially stabilizing positive definite terms have not been used, and terms involving gradients, e.g., $|\nablaperp \delta A|^2$,  could be estimated in conjunction  with those involving $(\delta A)^2$ by the Poincar\'{e} inequality, in order to obtain better results.  Also, the conditions we have obtained control $\delta A$ and $\delta f$, but an examination of $\delta^2 \mf_{MHD}$  reveals that there is a neutral direction given by $\delta \psi= \delta \Phi$.  This  was pointed out in \cite{MorEli}, where it was shown that this corresponds to the Alfv\'{e}n wave that has an interpretation in terms of spontaneous symmetry breaking and plays the role of the Goldstone mode of particle physics. 


\section{Conclusions}
\label{sec:conclu}

Stability analyses play a central role in the investigation of phenomena occurring in plasmas. For phenomena in which dissipative effects can be neglected, plasma models should be energy conserving and possess a Hamiltonian structure, thereby avoiding unphysical `phantom dissipation'.  For continuum models (e.g.,  kinetic or fluid theories) formulated in terms of Eulerian variables the Hamiltonian structure is generically of  noncanonical type.  Associated with such  Hamiltonian structure are Casimir functionals, particular invariants arising from  degeneracy of the cosymplectic (bivector) operator. The existence of Casimir invariants is the basis of the energy-Casimir method for determining stability conditions for noncanonical Hamiltonian systems which, as observed in Sec. \ref{sec:CS},  imply spectral stability.
In this paper we applied the energy-Casimir method to the planar version of a hybrid model for plasmas, which couples, via the pressure terms, the dynamics of a bulk MHD flow, with the kinetic evolution of a population of hot particles. 

After introducing the model, we formulated its Hamiltonian structure, pointing out how it  relates to  the Hamiltonian structures of reduced MHD and the Vlasov equation. In particular, in terms of the adapted variables, it emerged that the corresponding noncanonical Poisson bracket introduces the kinetic-MHD coupling terms, whereas the Hamiltonian is just the sum of the reduced MHD and Vlasov contributions. This was  reflected also in the Casimir structure. Indeed, the latter was seen to be decomposed into three independent contributions: two of these inherited from reduced MHD and Vlasov equation, which  correspond to the magnetic frozen-in condition and to the conservation of any function of $f$ integrated over phase space, respectively. However, the third family of Casimirs that was seen to  originate from the coupling terms in the bracket is peculiar to this model and expresses the conservation of a generalized hybrid cross-helicity, which, unlike the usual cross-helicity of MHD, accounts also for the contribution of the fluid momentum of the hot particle species. 

The abundance of Casimirs present in this planar reduction of the model, facilitated the application of the energy-Casimir method. From  the first variation of the free energy functional $\mf$, we determined general equations for equilibria of the system. These led, in particular, to a hybrid Grad-Shafranov equation, which generalizes the traditional equilibrium conditions of two-dimensional  MHD.
Finally, explicit energy stability conditions were obtained from the analysis of the second variation of $\mf$. On the basis of the obtained conditions, the presence of the hot particle species was seen to impose a lower bound on the equilibrium bulk speed, when compared to the pure MHD case. The presence of the kinetic component, also was seen to require stronger conditions on the current density profile for the stability to be attained. The distribution function, on the other hand, is constrained by dependence  on the MHD component, via the equilibrium relation, and its variation in terms of the equilibrium quantities required a  bounded from above in order to satisfy the stability conditions.  However, also in the presence of the hot particle population, we observed that without  MHD equilibrium flow  a monotonically decreasing current density profile satisfies the stability condition, as is the case for  reduced MHD.  Also, Gaussian distribution functions with unit variance are seen to satisfy the equilibrium condition, as is the case for purely kinetic Vlasov-like systems.

In closing, we remark that the conditions obtained are not optimal; further analysis of the  functional $\mf$ 
could lead to tighter conditions.  However, the energy-Casimir analysis as performed is direct and efficient, and circumvents more detailed spectral analysis.

\appendix


\section{Derivation of hybrid  Hamiltonian structure} 
\label{app1}
 
In this appendix we obtain the Hamiltonian structure, composed of the Hamiltonian of (\ref{ham}) and Poisson bracket of (\ref{planarbracket}), by restriction of the Hamiltonian structure  first given in Ref.~\cite{Tronci} for the full three-dimensional HPCS model. (For background material see, e.g., \cite{Mor98,Morrison2005} and see  \cite{AMP1} for a similar derivation.)

The HPCS Poisson bracket is given by
\begin{align}
\{F,G\}=& \int\!\de^3 x\, {\bM}\cdot\! \left(\frac{\delta G}{\delta
{\bM}}\cdot\nabla\frac{\delta
F}{\delta{\bM}}-\frac{\delta F}{\delta
{\bM}}\cdot\nabla\frac{\delta
G}{\delta{\bM}}\right)
 \label{PB-pressure-hybridMHD1}
 \\
&
-\int\!\de^3 x\,  \rho \left(\frac{\delta F}{\delta \bM}\cdot\nabla\frac{\delta G}{\delta \rho} 
-\frac{\delta G}{\delta \bM}\cdot \nabla\frac{\delta F}{\delta \rho}\right) 
\nonumber\\
&
+ \int\!\de^3 x\,  \bB\cdot\!\left(\frac{\delta F}{\delta \bM}\times\nabla\times\frac{\delta G}{\delta \bB}-\frac{\delta G}{\delta \bM}\times\nabla\times\frac{\delta F}{\delta \bB}\right) 
\nonumber
\\
&
+ \int\!\de^3 x \de^3 v\,  f\left(\left[\frac{\delta
F}{\delta f},\frac{\delta G}{\delta f}\right]_v
 +\, \bB\cdot\frac{\partial}{\partial \bv}\frac{\delta
F}{\delta f}
\times
\frac{\partial}{\partial \bv}\frac{\delta
G}{\delta f}
\right) 
\nonumber\\
&
+ \int\!\de^3 x \de^3 v\,  f\left(\left[\frac{\delta
F}{\delta f},\bv\cdot\frac{\delta G}{\delta \bM}\right]_v -\left[\frac{\delta
G}{\delta f},\bv\cdot\frac{\delta F}{\delta \bM}\right]_v\right)\,, 
\nonumber
\end{align}
whereas the Hamiltonian reads 
\begin{equation}
H=\int\!\de^3 x\, \left(\frac{|\bM|^2}{2\rho}\de^3 
 +  \rho\,\mathcal{U}(\rho) 
+ \frac{\left|\bB\right|^2}{2} \right)
+ \frac12\int\!\de^3 x \de^3 v  f \left|\bv\right|^2\,. 
\label{Ham-preMHD1}
\end{equation}
In Eqs.~(\ref{PB-pressure-hybridMHD1}) and (\ref{Ham-preMHD1}), $\rho$ and $\bM$ indicate the mass density of the bulk fluid and its momentum density, respectively, whereas $ \mathcal{U}(\rho)$ is the internal energy per unit mass. Physical constants have been set equal to unity.

The proof that (\ref{Ham-preMHD1}) satisfies the  Jacobi identity also can be carried out by  explicit verification.   Indeed,  upon recognizing that it is  composed of terms of the original bracket of  MHD \cite{MorrisonGreene} and that of the Maxwell-Vlasov system \cite{Morrison2bis, Morrison3,MaWe1,MaWeRaScSp}, together with later work on the two-fluid system \cite{Spencer,SpKa}, it is not difficult to ascertain the validity of the Jacobi identity.  Alternatively, one can begin with an action principle (see e.g., \cite{Morrison2}),  in particular the action principle for this model  of  \cite{HoTr2011}, and derive the Poisson bracket of (\ref{Ham-preMHD1}), thereby ensuring the Jacobi identity.

In order to see how the bracket (\ref{PB-pressure-hybridMHD1}) reduces to the bracket (\ref{planarbracket}) of the planar model, we consider first the two-dimensional restriction  of (\ref{PB-pressure-hybridMHD1}) by eliminating the dependence on the $z$ coordinate.  Then, we can enforce incompressibility by restricting to functionals that are independent on $\rho$  in (\ref{PB-pressure-hybridMHD1}). Consistently, we also remove the internal energy term from the Hamiltonian (\ref{Ham-preMHD1}). We remark that with such restrictions on the functionals and on the Poisson bracket, the Poisson bracket loses all explicit functional dependence on $\rho$.  Thus, according to the bracket theorem of \cite{Morrison3}, the restricted Poisson bracket must satisfy the Jacobi identity. 
 
By using vector identities, we can rewrite the first line of (\ref{PB-pressure-hybridMHD1}) as 
\begin{align}
&\int\de^2 x \, {\bM}\cdot\! \left(\frac{\delta G}{\delta
{\bM}}\cdot\nabla\frac{\delta
F}{\delta{\bM}}-\frac{\delta F}{\delta
{\bM}}\cdot\nabla\frac{\delta
G}{\delta{\bM}}\right)
\nonumber\\
&
= \int\de^2 x \left(\nabla\times{\bM}\cdot\! \left(\frac{\delta F}{\delta
{\bM}}\times\frac{\delta
G}{\delta{\bM}}\right)-\frac{\delta F}{\delta
{\bM}}\nabla\cdot\frac{\delta
G}{\delta{\bM}}+\frac{\delta G}{\delta
{\bM}}\nabla\cdot\frac{\delta
F}{\delta{\bM}}\right)\,.
 \label{PB2d}
 \end{align}
Then, introduction of  the relations
\begin{equation} \label{chg}
\nabla\times\bM=\omega \hat{z}, \qquad \bB=\nabla\times(A\hat{z}),
\end{equation}
leads to the following rule for transforming the functional derivatives
\begin{equation} \label{rulefd}
\frac{\delta F}{\delta \bM}=\nabla\times\left(\frac{\delta \bar{F}}{\delta \omega}\hat{z}\right), \qquad \hat{z}\cdot\nabla\times \frac{\delta F}{\delta \bB}=\frac{\delta \bar{F}}{\delta A}.
\end{equation}
Using (\ref{chg})-(\ref{rulefd}), together with (\ref{PB2d}) in (\ref{PB-pressure-hybridMHD1}), leads namely to the  bracket (\ref{planarbracket}) of the incompressible planar model. The Hamiltonian (\ref{Ham-preMHD1}), on the other hand, reduces to \eqref{ham}.


\section{Casimir verification}
\label{app2}

Here we demonstrate explicitly that  \eqref{NewCasimir} is a Casimir invariant, i.e., satisfies 
 $\{F,C\}=0$ for all functionals $F$.  This is simplified by noting  that  $\int \de^2x \, \Psi(A)$ and $\int \de^2x\,  \de^3 v\Lambda(f)$ are separately  Casimirs of   \eqref{planarbracket}.  Using  $\delta{C}/\delta{f}=\bv\cdot\hod\Phi$, we obtain 
\begin{align*}
\{F,C\}=&\int \de^2 x \, \omega\left[\dede{F}{\omega},\Phi\right] 
+\int\! \de^2 x \, A\left(\left[\dede{F}{\omega},\overline{\omega}\,\Phi'\right]-\left[\Phi,\dede{F}{A}\right]\right)
\\
&\hspace{1 cm} -\int \de^2 x \de^3 v\,  f \left(\left[\dede{F}{f},\bv\cdot\hod\Phi\right]_v 
-\left[\bv\cdot\hod\Phi,\bv\cdot\hod\dede{F}{\omega}\right]_v\right)
\\
&\hspace{2 cm}+\int\! \de^2 x \de^3 v \,  f\, \left[\dede{F}{f},\bv\cdot\hod\Phi \right]_v
\\
=&\int \de^2 x\,  \omega\left[\dede{F}{\omega},\Phi\right]
- \int\!  \de^2 x \, \overline{\omega}\left[\dede{F}{\omega},\Phi\right]
\\
& \hspace{1 cm} +
\int\! \de^2 x \de^3 v\, f\left[\bv\cdot\hod\Phi,\bv\cdot\hod\dede{F}{\omega}\right]_v
\\
=&
\int\! \de^2 x \de^3 v \,   \bv\cdot\hod f\left[\dede{F}{\omega},\Phi\right]
\\
&\hspace{1 cm} -\int\! \de^2 x \de^3 v  \,  f\left(\left[\bv\cdot\hod\Phi,\dede{F}{\omega}\right]-\left[\bv\cdot\hod\dede{F}{\omega},\Phi\right]\right)
\\
=&
\int\! \de^2 x \de^3 v \,  \bv\cdot\hod f\left[\dede{F}{\omega},\Phi\right]
-\int\! \de^2 x \de^3 v \,  f\left(\bv\cdot\hod\left[\Phi,\dede{F}{\omega}\right]\right.
\\
& \hspace{2 cm}  \left. -\left[\Phi,\bv\cdot\hod\dede{F}{\omega}\right]-\left[\bv\cdot\hod\dede{F}{\omega},\Phi\right]\right)
= 0
\,.
\end{align*}
In the above computations, integrations by parts with vanishing boundary terms have been carried out and the Leibniz identity has been used. Also,   prime denotes differentiation  with respect to the argument.

\section*{Acknowledgements.}

\noindent PJM was supported by US Department of Energy, Grant No.~DE-FG02-04ER54742. This work was supported by the European Community under the contracts of Association between EURATOM, CEA, and the French Research Federation for fusion studies. The views and opinions expressed herein do not necessarily reflect those of the European Commission. ET received financial support from the Agence Nationale de la Recherche (ANR GYPSI) and from the CNRS (PEPS project GEOPLASMA).

\bibliographystyle{apsrev}

\begin{thebibliography}{33}
\expandafter\ifx\csname natexlab\endcsname\relax\def\natexlab#1{#1}\fi
\expandafter\ifx\csname bibnamefont\endcsname\relax
  \def\bibnamefont#1{#1}\fi
\expandafter\ifx\csname bibfnamefont\endcsname\relax
  \def\bibfnamefont#1{#1}\fi
\expandafter\ifx\csname citenamefont\endcsname\relax
  \def\citenamefont#1{#1}\fi
\expandafter\ifx\csname url\endcsname\relax
  \def\url#1{\texttt{#1}}\fi
\expandafter\ifx\csname urlprefix\endcsname\relax\def\urlprefix{URL }\fi
\providecommand{\bibinfo}[2]{#2}
\providecommand{\eprint}[2][]{\url{#2}}

\bibitem[{\citenamefont{Holm et~al.}(1985)\citenamefont{Holm, Marsden, Ratiu,
  and Weinstein}}]{HoMaRaWe}
\bibinfo{author}{\bibfnamefont{D.~D.} \bibnamefont{Holm}},
  \bibinfo{author}{\bibfnamefont{J.~E.} \bibnamefont{Marsden}},
  \bibinfo{author}{\bibfnamefont{T.~S.} \bibnamefont{Ratiu}}, \bibnamefont{and}
  \bibinfo{author}{\bibfnamefont{A.}~\bibnamefont{Weinstein}},
  \bibinfo{journal}{Phys. Rep.} \textbf{\bibinfo{volume}{123}},
  \bibinfo{pages}{1} (\bibinfo{year}{1985}).

\bibitem[{\citenamefont{Morrison}(1998)}]{Mor98}
\bibinfo{author}{\bibfnamefont{P.~J.} \bibnamefont{Morrison}},
  \bibinfo{journal}{Rev. Mod. Phys.} \textbf{\bibinfo{volume}{70}},
  \bibinfo{pages}{467} (\bibinfo{year}{1998}).

\bibitem[{\citenamefont{Yoshida et~al.}(2013)\citenamefont{Yoshida, Morrison,
  and Dobarro}}]{yoshida}
\bibinfo{author}{\bibfnamefont{Z.}~\bibnamefont{Yoshida}},
  \bibinfo{author}{\bibfnamefont{P.~J.} \bibnamefont{Morrison}},
  \bibnamefont{and} \bibinfo{author}{\bibfnamefont{F.}~\bibnamefont{Dobarro}},
  \bibinfo{journal}{J. Math. Fluid Mech.} \textbf{\bibinfo{volume}{accepted}}
  (\bibinfo{year}{2013}).

\bibitem[{\citenamefont{Hazeltine et~al.}(1984)\citenamefont{Hazeltine, Holm,
  E., and Morrison}}]{Haz84}
\bibinfo{author}{\bibfnamefont{R.~D.} \bibnamefont{Hazeltine}},
  \bibinfo{author}{\bibfnamefont{D.~D.} \bibnamefont{Holm}},
  \bibinfo{author}{\bibfnamefont{M.~J.} \bibnamefont{E.}}, \bibnamefont{and}
  \bibinfo{author}{\bibfnamefont{P.~J.} \bibnamefont{Morrison}}, in
  \emph{\bibinfo{booktitle}{International Conference on Plasma Physics
  Proceedings}}, edited by \bibinfo{editor}{\bibfnamefont{M.~Q.}
  \bibnamefont{Tran}} \bibnamefont{and} \bibinfo{editor}{\bibfnamefont{M.~L.}
  \bibnamefont{Sawley}} (\bibinfo{organization}{\'Ecole Polytechinque
  F\'ed\'erale de Lausanne}, \bibinfo{address}{Lausanne},
  \bibinfo{year}{1984}), pp. \bibinfo{pages}{203--211}.

\bibitem[{\citenamefont{Morrison and Eliezer}(1986)}]{MorEli}
\bibinfo{author}{\bibfnamefont{P.~J.} \bibnamefont{Morrison}} \bibnamefont{and}
  \bibinfo{author}{\bibfnamefont{S.}~\bibnamefont{Eliezer}},
  \bibinfo{journal}{Phys. Rev. A} \textbf{\bibinfo{volume}{33}},
  \bibinfo{pages}{4205} (\bibinfo{year}{1986}).

\bibitem[{\citenamefont{Batt et~al.}(1995)\citenamefont{Batt, Morrison, and
  Rein}}]{Batt}
\bibinfo{author}{\bibfnamefont{J.}~\bibnamefont{Batt}},
  \bibinfo{author}{\bibfnamefont{P.~J.} \bibnamefont{Morrison}},
  \bibnamefont{and} \bibinfo{author}{\bibfnamefont{G.}~\bibnamefont{Rein}},
  \bibinfo{journal}{Arch. Rat. Mech. Anal.} \textbf{\bibinfo{volume}{130}},
  \bibinfo{pages}{163} (\bibinfo{year}{1995}).

\bibitem[{\citenamefont{Rein}(1994)}]{Rein}
\bibinfo{author}{\bibfnamefont{G.}~\bibnamefont{Rein}}, \bibinfo{journal}{Math.
  Meth. Appl. Sci.} \textbf{\bibinfo{volume}{17}}, \bibinfo{pages}{1129}
  (\bibinfo{year}{1994}).

\bibitem[{\citenamefont{Cheng}(1991)}]{Cheng}
\bibinfo{author}{\bibfnamefont{C.~Z.} \bibnamefont{Cheng}},
  \bibinfo{journal}{J. Geophys. Res.} \textbf{\bibinfo{volume}{96}},
  \bibinfo{pages}{21159} (\bibinfo{year}{1991}).

\bibitem[{\citenamefont{Park and et~al.}(1992)}]{ParkEtAl}
\bibinfo{author}{\bibfnamefont{W.}~\bibnamefont{Park}} \bibnamefont{and}
  \bibinfo{author}{\bibnamefont{et~al.}}, \bibinfo{journal}{Phys. Fluids B}
  \textbf{\bibinfo{volume}{4}}, \bibinfo{pages}{2033} (\bibinfo{year}{1992}).

\bibitem[{\citenamefont{Park et~al.}(1999)\citenamefont{Park, Belova, Fu, Tang,
  Strauss, and Sugiyama}}]{PaBeFuTaStSu}
\bibinfo{author}{\bibfnamefont{W.}~\bibnamefont{Park}},
  \bibinfo{author}{\bibfnamefont{E.~V.} \bibnamefont{Belova}},
  \bibinfo{author}{\bibfnamefont{G.~Y.} \bibnamefont{Fu}},
  \bibinfo{author}{\bibfnamefont{X.~Z.} \bibnamefont{Tang}},
  \bibinfo{author}{\bibfnamefont{H.~R.} \bibnamefont{Strauss}},
  \bibnamefont{and} \bibinfo{author}{\bibfnamefont{L.}~\bibnamefont{Sugiyama}},
  \bibinfo{journal}{Phys. Plasmas} \textbf{\bibinfo{volume}{6}},
  \bibinfo{pages}{1796} (\bibinfo{year}{1999}).

\bibitem[{\citenamefont{Fu and Park}(1995)}]{FuPark}
\bibinfo{author}{\bibfnamefont{G.~Y.} \bibnamefont{Fu}} \bibnamefont{and}
  \bibinfo{author}{\bibfnamefont{W.}~\bibnamefont{Park}},
  \bibinfo{journal}{Phys. Rev. Lett.} \textbf{\bibinfo{volume}{74}},
  \bibinfo{pages}{1594} (\bibinfo{year}{1995}).

\bibitem[{\citenamefont{Kim et~al.}(2004)\citenamefont{Kim, Sovinec, and
  Parker}}]{KimEtAl}
\bibinfo{author}{\bibfnamefont{C.~C.} \bibnamefont{Kim}},
  \bibinfo{author}{\bibfnamefont{C.~R.} \bibnamefont{Sovinec}},
  \bibnamefont{and} \bibinfo{author}{\bibfnamefont{S.~E.}
  \bibnamefont{Parker}}, \bibinfo{journal}{Comp. Phys. Comm.}
  \textbf{\bibinfo{volume}{164}}, \bibinfo{pages}{448} (\bibinfo{year}{2004}).

\bibitem[{\citenamefont{Takahash et~al.}(2009)\citenamefont{Takahash, Brennan,
  and Kim}}]{TaBrKi}
\bibinfo{author}{\bibfnamefont{R.}~\bibnamefont{Takahash}},
  \bibinfo{author}{\bibfnamefont{D.~P.} \bibnamefont{Brennan}},
  \bibnamefont{and} \bibinfo{author}{\bibfnamefont{C.~C.} \bibnamefont{Kim}},
  \bibinfo{journal}{Phys. Rev. Lett.} \textbf{\bibinfo{volume}{102}},
  \bibinfo{pages}{135001} (\bibinfo{year}{2009}).

\bibitem[{\citenamefont{Holm and Tronci}(2012)}]{HoTr2011}
\bibinfo{author}{\bibfnamefont{D.~D.} \bibnamefont{Holm}} \bibnamefont{and}
  \bibinfo{author}{\bibfnamefont{C.}~\bibnamefont{Tronci}},
  \bibinfo{journal}{Comm. Math. Sci.} \textbf{\bibinfo{volume}{10}},
  \bibinfo{pages}{191} (\bibinfo{year}{2012}).

\bibitem[{\citenamefont{Tronci}(2010)}]{Tronci}
\bibinfo{author}{\bibfnamefont{C.}~\bibnamefont{Tronci}}, \bibinfo{journal}{J.
  Phys. A: Math. Theor.} \textbf{\bibinfo{volume}{43}}, \bibinfo{pages}{375501}
  (\bibinfo{year}{2010}).

\bibitem[{\citenamefont{Tronci et~al.}(2013)\citenamefont{Tronci, Tassi, and
  Morrison}}]{TrTaMo}
\bibinfo{author}{\bibfnamefont{C.}~\bibnamefont{Tronci}},
  \bibinfo{author}{\bibfnamefont{E.}~\bibnamefont{Tassi}}, \bibnamefont{and}
  \bibinfo{author}{\bibfnamefont{P.~J.} \bibnamefont{Morrison}},
  \bibinfo{journal}{preprint}  (\bibinfo{year}{2013}).

\bibitem[{\citenamefont{Morrison and Hazeltine}(1984)}]{Mor84}
\bibinfo{author}{\bibfnamefont{P.~J.} \bibnamefont{Morrison}} \bibnamefont{and}
  \bibinfo{author}{\bibfnamefont{R.~D.} \bibnamefont{Hazeltine}},
  \bibinfo{journal}{Phys. Fluids} \textbf{\bibinfo{volume}{27}},
  \bibinfo{pages}{886} (\bibinfo{year}{1984}).

\bibitem[{\citenamefont{Marsden and Morrison}(1984)}]{MaMo}
\bibinfo{author}{\bibfnamefont{J.~E.} \bibnamefont{Marsden}} \bibnamefont{and}
  \bibinfo{author}{\bibfnamefont{P.~J.} \bibnamefont{Morrison}},
  \bibinfo{journal}{Contemp. Math.} \textbf{\bibinfo{volume}{28}},
  \bibinfo{pages}{133} (\bibinfo{year}{1984}).

\bibitem[{\citenamefont{Morrison}(1980)}]{Morrison2bis}
\bibinfo{author}{\bibfnamefont{P.~J.} \bibnamefont{Morrison}},
  \bibinfo{journal}{Phys. Lett. A} \textbf{\bibinfo{volume}{80}},
  \bibinfo{pages}{383} (\bibinfo{year}{1980}).

\bibitem[{\citenamefont{Morrison}(1982)}]{Morrison3}
\bibinfo{author}{\bibfnamefont{P.~J.} \bibnamefont{Morrison}}, in
  \emph{\bibinfo{booktitle}{Mathematical Methods in Hydrodynamics and
  Integrability in Dynamical Systems}}, edited by
  \bibinfo{editor}{\bibfnamefont{M.}~\bibnamefont{Tabor}} \bibnamefont{and}
  \bibinfo{editor}{\bibfnamefont{Y.}~\bibnamefont{Treve}}
  (\bibinfo{publisher}{AIP Conference Proceedings}, \bibinfo{year}{1982}), pp.
  \bibinfo{pages}{13--46}.

\bibitem[{\citenamefont{Marsden and Weinstein}(1982)}]{MaWe1}
\bibinfo{author}{\bibfnamefont{J.~E.} \bibnamefont{Marsden}} \bibnamefont{and}
  \bibinfo{author}{\bibfnamefont{A.}~\bibnamefont{Weinstein}},
  \bibinfo{journal}{Phys. D} \textbf{\bibinfo{volume}{4}}, \bibinfo{pages}{394}
  (\bibinfo{year}{1982}).

\bibitem[{\citenamefont{Morrison}(1987)}]{Mor87}
\bibinfo{author}{\bibfnamefont{P.~J.} \bibnamefont{Morrison}},
  \bibinfo{journal}{Zeitschrift f\"ur Naturforschung}
  \textbf{\bibinfo{volume}{42a}}, \bibinfo{pages}{1115} (\bibinfo{year}{1987}).

\bibitem[{\citenamefont{Fowler}(1963)}]{Fowler}
\bibinfo{author}{\bibfnamefont{K.}~\bibnamefont{Fowler}}, \bibinfo{journal}{J.
  Math. Phys.} \textbf{\bibinfo{volume}{4}}, \bibinfo{pages}{559}
  (\bibinfo{year}{1963}).

\bibitem[{\citenamefont{Gardner}(1963)}]{Gar63}
\bibinfo{author}{\bibfnamefont{C.~S.} \bibnamefont{Gardner}},
  \bibinfo{journal}{Phys. Fluids} \textbf{\bibinfo{volume}{6}},
  \bibinfo{pages}{839} (\bibinfo{year}{1963}).

\bibitem[{\citenamefont{Morrison and Kotschenreuther}(1990)}]{Mor90}
\bibinfo{author}{\bibfnamefont{P.~J.} \bibnamefont{Morrison}} \bibnamefont{and}
  \bibinfo{author}{\bibfnamefont{M.}~\bibnamefont{Kotschenreuther}}, in
  \emph{\bibinfo{booktitle}{Nonlinear World. IV International Workshop on
  Nonlinear and Turbulent Processes in Physics}}, edited by
  \bibinfo{editor}{\bibfnamefont{V.~G.} \bibnamefont{Bar'yakhtar}},
  \bibinfo{editor}{\bibfnamefont{V.~M.} \bibnamefont{Chernousenko}},
  \bibinfo{editor}{\bibfnamefont{N.~S.} \bibnamefont{Erokhin}},
  \bibinfo{editor}{\bibfnamefont{A.~B.} \bibnamefont{Sitenko}},
  \bibnamefont{and} \bibinfo{editor}{\bibfnamefont{V.~E.}
  \bibnamefont{Zakharov}} (\bibinfo{publisher}{World Scientific},
  \bibinfo{address}{Singapore}, \bibinfo{year}{1990}), pp.
  \bibinfo{pages}{910--932}.

\bibitem[{\citenamefont{Morrison et~al.}(2013)\citenamefont{Morrison, Tassi,
  and Tronko}}]{Mor13}
\bibinfo{author}{\bibfnamefont{P.~J.} \bibnamefont{Morrison}},
  \bibinfo{author}{\bibfnamefont{E.}~\bibnamefont{Tassi}}, \bibnamefont{and}
  \bibinfo{author}{\bibfnamefont{N.}~\bibnamefont{Tronko}},
  \bibinfo{journal}{Phys. Plasmas} \textbf{\bibinfo{volume}{20}},
  \bibinfo{pages}{042109} (\bibinfo{year}{2013}).

\bibitem[{\citenamefont{Morrison}(2005)}]{Morrison2005}
\bibinfo{author}{\bibfnamefont{P.~J.} \bibnamefont{Morrison}},
  \bibinfo{journal}{Phys. Plasmas} \textbf{\bibinfo{volume}{12}},
  \bibinfo{pages}{058102} (\bibinfo{year}{2005}).

\bibitem[{\citenamefont{Andreussi et~al.}(2012)\citenamefont{Andreussi,
  Morrison, and Pegoraro}}]{AMP1}
\bibinfo{author}{\bibfnamefont{T.}~\bibnamefont{Andreussi}},
  \bibinfo{author}{\bibfnamefont{P.~J.} \bibnamefont{Morrison}},
  \bibnamefont{and} \bibinfo{author}{\bibfnamefont{F.}~\bibnamefont{Pegoraro}},
  \bibinfo{journal}{Phys. Plasmas} \textbf{\bibinfo{volume}{19}},
  \bibinfo{pages}{052102} (\bibinfo{year}{2012}).

\bibitem[{\citenamefont{Morrison and Greene}(1980)}]{MorrisonGreene}
\bibinfo{author}{\bibfnamefont{P.~J.} \bibnamefont{Morrison}} \bibnamefont{and}
  \bibinfo{author}{\bibfnamefont{J.~M.} \bibnamefont{Greene}},
  \bibinfo{journal}{Phys. Rev. Lett.} \textbf{\bibinfo{volume}{45}},
  \bibinfo{pages}{790} (\bibinfo{year}{1980}).

\bibitem[{\citenamefont{Marsden et~al.}(1983)\citenamefont{Marsden, Weinstein,
  Ratiu, Schmid, and Spencer}}]{MaWeRaScSp}
\bibinfo{author}{\bibfnamefont{J.~E.} \bibnamefont{Marsden}},
  \bibinfo{author}{\bibfnamefont{A.}~\bibnamefont{Weinstein}},
  \bibinfo{author}{\bibfnamefont{T.~S.} \bibnamefont{Ratiu}},
  \bibinfo{author}{\bibfnamefont{R.}~\bibnamefont{Schmid}}, \bibnamefont{and}
  \bibinfo{author}{\bibfnamefont{R.~G.} \bibnamefont{Spencer}},
  \bibinfo{journal}{Atti Accad. Sci. Torino Cl. Sci. Fis. Mat. Natur.}
  \textbf{\bibinfo{volume}{117}}, \bibinfo{pages}{289} (\bibinfo{year}{1983}).

\bibitem[{\citenamefont{Spencer}(1982)}]{Spencer}
\bibinfo{author}{\bibfnamefont{R.~G.} \bibnamefont{Spencer}}, in
  \emph{\bibinfo{booktitle}{Mathematical Methods in Hydrodynamics and
  Integrability in Dynamical Systems}}, edited by
  \bibinfo{editor}{\bibfnamefont{M.}~\bibnamefont{Tabor}} \bibnamefont{and}
  \bibinfo{editor}{\bibfnamefont{Y.}~\bibnamefont{Treve}}
  (\bibinfo{publisher}{AIP Conference Proceedings}, \bibinfo{address}{La
  Jolla}, \bibinfo{year}{1982}), pp. \bibinfo{pages}{121--126}.

\bibitem[{\citenamefont{Spencer and Kaufman}(1982)}]{SpKa}
\bibinfo{author}{\bibfnamefont{R.~G.} \bibnamefont{Spencer}} \bibnamefont{and}
  \bibinfo{author}{\bibfnamefont{A.~N.} \bibnamefont{Kaufman}},
  \bibinfo{journal}{Phys. Rev. A} \textbf{\bibinfo{volume}{25}},
  \bibinfo{pages}{2437} (\bibinfo{year}{1982}).

\bibitem[{\citenamefont{Morrison}(2009)}]{Morrison2}
\bibinfo{author}{\bibfnamefont{P.~J.} \bibnamefont{Morrison}}, in
  \emph{\bibinfo{booktitle}{New developments in Nonlinear Plasma Physics:
  Proceedings for the 2009 ICTP College on Plasma Physics}}, edited by
  \bibinfo{editor}{\bibfnamefont{B.}~\bibnamefont{Eliasson}} \bibnamefont{and}
  \bibinfo{editor}{\bibfnamefont{P.}~\bibnamefont{Shukla}}
  (\bibinfo{publisher}{AIP Conference Proceedings}, \bibinfo{address}{La
  Jolla}, \bibinfo{year}{2009}), pp. \bibinfo{pages}{329--344}.

\end{thebibliography}

\end{document}